# Emergence of Finite-Time Singularities from Accelerated Event Recurrence: Insights into the Mechanism of Catastrophic Failure

Qinghua Lei[1], Didier Sornette[2]

[1]Department of Earth Sciences, Uppsala University, Sweden

[2]Institute of Risk Analysis, Prediction and Management, Academy for Advanced Interdisciplinary Studies, Southern University of Science and Technology, Shenzhen, China

Correspondence: Qinghua Lei (qinghua.lei@geo.uu.se) and Didier Sornette (dsornette@ethz.ch)

Qinghua Lei and Didier Sornette contributed equally to this work.

**Key Points:**

- We develop discrete-event models capturing how damage accumulation leads to finite-time singularities before failure.
- Logarithmic and power law singularities emerge from shrinking interevent times and growing event magnitudes.
- A mean-field model links power law exponents to system stiffness, explaining diverse precursory acceleration patterns.

**Abstract**

We develop a discrete-event modeling framework that captures the progression of geophysical systems toward catastrophic failure through sequences of distinct damage events. By representing system evolution as a succession of temporally accelerating and amplitude-varying events, the framework reveals how finite-time singularities, both logarithmic and power law types, naturally emerge from the interplay between shrinking interevent intervals and growing event magnitudes. This event-based perspective provides an intuitive physical understanding of rupture processes, highlighting how precursory signals such as accelerating strain rate, event frequency, and energy release can be traced back to simple underlying mechanisms. A mean-field formulation further links the observed power law exponents to the evolving stiffness of the system under constant or time-varying stress. Incorporating stochastic fluctuations, the model captures the inherent randomness of natural systems leading to the emergence of stochastic finite-time singular behavior. Altogether, this unified approach offers a simple conceptual and quantitative tool for interpreting the lead-up to failure in a wide range of geophysical settings.

**Plain Language Summary**

We present a new way to understand how natural systems, such as landslides, glaciers, and volcanoes, progress toward catastrophic failure. Rather than treating these processes as smooth and continuous, we model them as a sequence of discrete damage events that punctuate the system's evolution over time. This event-based perspective, similar to how quantum physics describes light as discrete quanta rather than a continuous wave, provides a more intuitive understanding of how failure unfolds in complex natural systems. Our models show that, as these discrete events occur more frequently and with increasing intensity, the system may approach a so-called "finite-time singularity"—a mathematical signature of an impending rupture or collapse. This behavior helps explain why warning signals such as accelerating ground movement, increasing rates of small precursory events, and growing energy release are often observed before major geophysical disasters. We also examine how natural variability and randomness influence these patterns, showing that, even with stochasticity, the approach to failure can exhibit predictable statistical features. Overall, this framework offers both a conceptual and quantitative tool for interpreting warning signs, with potential to improve early warning systems and our broader understanding of extreme geophysical hazards.

## 1 Introduction

Catastrophic failure occurs in diverse geomaterials, including rock, soil, and ice, and drives a broad range of geophysical hazards, such as landslides, glacier breakoffs, and volcanic eruptions (Caricchi et al., 2021; Faillettaz et al., 2015; Lacroix et al., 2020; Lei & Sornette, 2025d). Advancing our fundamental understanding of the underlying mechanisms is essential for improving our ability to predict those extreme geohazard events.

One mechanism for explaining catastrophic failure is based on an analogy with critical phase transitions, where the system evolves toward a critical point characterized by scaling symmetry and a finite-time singularity (Ausloos, 1986; Main, 1999; Johansen & Sornette, 2000; Sammis & Sornette, 2002; Sornette, 2002). The scaling symmetry implies that the system would exhibit power law behavior as it approaches catastrophic failure, consistent with frequently observed precursory power law dynamics in landslides, earthquakes, rockbursts, glaciers, and volcanoes (Amitrano et al., 2005; Bufe & Varnes, 1993; Faillettaz et al., 2015; Intrieri et al., 2019; Jaumé & Sykes, 1999; Lei & Sornette, 2025d; Ouillon & Sornette, 2000; Pralong & Funk, 2005; Voight, 1988, 1989). The presence of a finite-time singularity reflects the system's transition near the critical point between distinct physical regimes—specifically, from a quasi-static deformation regime driven by progressive damage accumulation to an elasto-dynamic rupture regime governed by the finite speed of elastic waves (Sammis & Sornette, 2002). Extensive studies have also investigated the micromechanical origins of power law acceleration, attributing it to processes such as subcritical crack growth driven by stress corrosion (Cornelius & Scott, 1993; Heap et al., 2009; Kilburn, 2012; Kilburn & Voight, 1998; Main, 2000; Sammis & Sornette, 2002) and rate- and state-dependent friction along sliding surfaces (Helmstetter et al., 2004; Paul et al., 2024; Noda & Chang, 2023). While those studies offer valuable insights into the physical origins of accelerating precursors, they remain limited in explaining the diverse patterns of observational time series data at the site scale.

In this paper, we develop a suite of reduced event-based models grounded in a first-principle assumption that catastrophic failure arises from accelerating occurrence of damage events. Our results provide broadly applicable insights into the mechanism underlying finite-time singularities observed in diverse geophysical systems. The remainder of the paper is organized as follows. Section 2 introduces the mathematical formulation of finite-time singularities and provides illustrative examples from diverse geophysical hazards. Section 3 presents the derivation of a series of reduced event-based models, accompanied by illustrative calculation examples. Section 4 discusses the underlying mechanisms of catastrophic failure and explores the relationships between finite-time singularities captured by different observable quantities, with a conclusion provided at the end.

## 2 Finite-time singularity

The evolution of a geophysical system, such as a landslide, glacier, or volcano, during the terminal stage preceding catastrophic failure is generically governed by the following nonlinear dynamical equation (Voight 1988, 1989; see Intrieri *et al.* 2019 for a review of related methods and their history):

$$\ddot{\Omega} = \eta \dot{\Omega}^\alpha, \text{ with } \alpha > 1, \quad (1)$$

where $\Omega$ is an observable quantity (e.g., displacement, strain, tilt, energy release, and earthquake count) with the overdots denoting time derivatives, $\eta$ is a constant, and $\alpha$ is an exponent characterizing the degree of nonlinearity. The condition $\alpha > 1$ signifies the existence of positive feedbacks, leading to a super-exponential dynamic that culminates in a finite-time singularity (Ide & Sornette, 2002; Lei & Sornette, 2023, 2025d; Main, 1999; Sammis & Sornette, 2002). This singular behavior becomes evident by integrating equation (1), which gives:

$$\dot{\Omega} = \frac{\kappa}{(t_c - t)^p}, \text{ with } p > 0, \quad (2)$$

where $\kappa = (p/\eta)^p$ is a constant, $t$ is time, $p = 1/(\alpha-1)$ is an exponent with the condition $p > 0$ ensuring the existence of a finite-time singularity at the critical time $t_c$, around which the system transitions from a quasi-static regime (driven by progressive damage) to an elasto-dynamical regime (characterized by runaway rupture). A further integration of equation (2) leads to:

$$\Omega(t) = \begin{cases} A - \dfrac{\kappa}{m}(t_c - t)^m, & m \neq 0 \\ A - \kappa \ln(t_c - t), & m = 0 \end{cases}, \quad (3)$$

where $m = 1-p = (\alpha-2)/(\alpha-1) < 1$ is a critical exponent reflecting the strength of the finite-time singularity and $A$ is a constant determined from the initial condition of $\Omega(t = t_0) = \Omega_0$. The case $m = 0$ corresponds to $p = 1$ and $\alpha = 2$. We rewrite equation (3) as:

$$\Omega(t) = \begin{cases} \Omega_0 + \dfrac{\kappa}{m}\left[(t_c - t_0)^m - (t_c - t)^m\right], & m \neq 0 \\ \Omega_0 - \kappa \ln\left(\dfrac{t_c - t}{t_c - t_0}\right), & m = 0 \end{cases}. \quad (4)$$

Note that the solution of $\Omega(t)$ for $m \neq 0$ (with a power law finite-time singularity) converges to the solution for $m = 0$ (with a logarithmic finite-time singularity) as $m \to 0$. This is seen by applying the Taylor expansion $(t_c-t_0)^m = \exp[m\ln(t_c-t_0)] = 1+m\ln(t_c-t_0)+O(m^2)$ and $(t_c-t)^m = \exp[m\ln(t_c-t)] = 1+m\ln(t_c-t)+O(m^2)$, which obtains:

$$\Omega(t) = \Omega_0 + \frac{\kappa}{m}\left[e^{m\ln(t_c-t_0)} - e^{m\ln(t_c-t)}\right] = \Omega_0 - \kappa \ln\left(\frac{t_c - t}{t_c - t_0}\right) + O(m). \quad (5)$$

In other words, the logarithmic finite-time singularity for $m = 0$ and $p = 1$ is a special case of the general power law finite-time singularity.

By defining $B = -\kappa/m$ and substituting it into equation (3), we can finally obtain the general solution of $\Omega(t)$ for $m < 1$ (including $m = 0$):

$$\Omega(t) = A + B(t_c - t)^m, \text{ with } m < 1, \tag{6}$$

expressing that the observable $\Omega$ is a power law function of the time to failure $t_c$–$t$. For $0 < m < 1$, $\dot{\Omega}$ diverges at $t_c$, while $\Omega$ converges to the finite value $A$; for $m < 0$, both $\dot{\Omega}$ and $\Omega$ diverge at $t_c$. Of course, physically, no divergence ever occurs. The existence of the mathematical divergence indicates a transition to the elasto-dynamical rupture regime that occurs close to $t_c$ with the rupture accelerating very rapidly, constrained only by the speed of elastic waves. The important insight is that, for a large part of the evolution of $\Omega$, its dynamics is accurately described by equation (6), as if a genuine finite-time singularity would occur. It is only close to $t_c$ that a change of regime leads to a new elasto-dynamical phase ending in rupture.

The power law finite-time singularity formation, i.e., equation (6), can describe the precursory acceleration behavior observed in many natural systems, such as landslides, glaciers, and volcanoes before catastrophic events. Figure 1 gives some typical examples, including the Brienz/Brinzauls landslide in Switzerland (Loew et al., 2024), the Preonzo landslide in Switzerland (Loew et al., 2017), the Abbotsford landslide in New Zealand (Hancox, 2008), the Grandes Jorasses glacier in Italy (Faillettaz et al., 2015), the Weissmies glacier in Switzerland (Meier et al., 2018), the Amery iceshelf in Antarctica (Walker et al., 2021), the Sierra Negra volcano in Ecuador (Chadwick et al., 2006), the St Helens volcano in USA (Dzurisin et al., 1983), and Merapi volcano in Indonesia (Surono et al., 2012). Here, we fit equation (6) to various cumulative observables, including displacement, tilt angle, rift length, and earthquake count, using a stable and robust calibration scheme (Lei & Sornette, 2025d). The resulting model parameters are then used to derive the parameters in equation (2), which is subsequently compared with the corresponding rate measurements (see insets in Figure 1).

In general, equation (6) captures well the overall acceleration trend across all examples. In contrast, the corresponding rate data sometimes display pronounced fluctuations around the prediction of equation (2), likely due to the influence of high-frequency noise that dominates rate measurements. These fluctuations are much less important in the cumulative observables, as integration inherently acts as a low-pass filter that suppresses high-frequency noise (Huang et al., 2000). Note that some of the fluctuations around the power law trend may also reflect underlying log-periodic oscillations, which arise from intermittent rupture dynamics in heterogeneous geophysical systems (Faillettaz et al., 2008; Lei & Sornette, 2025d, 2025b, 2025c, 2025a). Some deviations observed in the rate data near the final failure time may also be attributed to the emergence of dragon-king events (a double metaphor for an event of a predominant impact/size like a "king" and a unique origin like a "dragon") (Sornette & Ouillon, 2012; Lei et al., 2023), which reflect the amplification of the deformation process via positive feedback mechanisms as the system approaches catastrophic failure.

In the present work, we focus on the general power law acceleration trend. To this end, we construct a family of models in which system evolution is represented as a succession of damage events occurring with accelerating frequency, with the goal of identifying a reduced-form framework that captures, in simple and intuitive terms, the physical mechanisms leading to finite-time singularities. By focusing on the interplay between accelerating event rates and amplitude growth, we seek to establish a transparent mechanism that captures how such singularities emerge naturally across diverse geophysical systems.

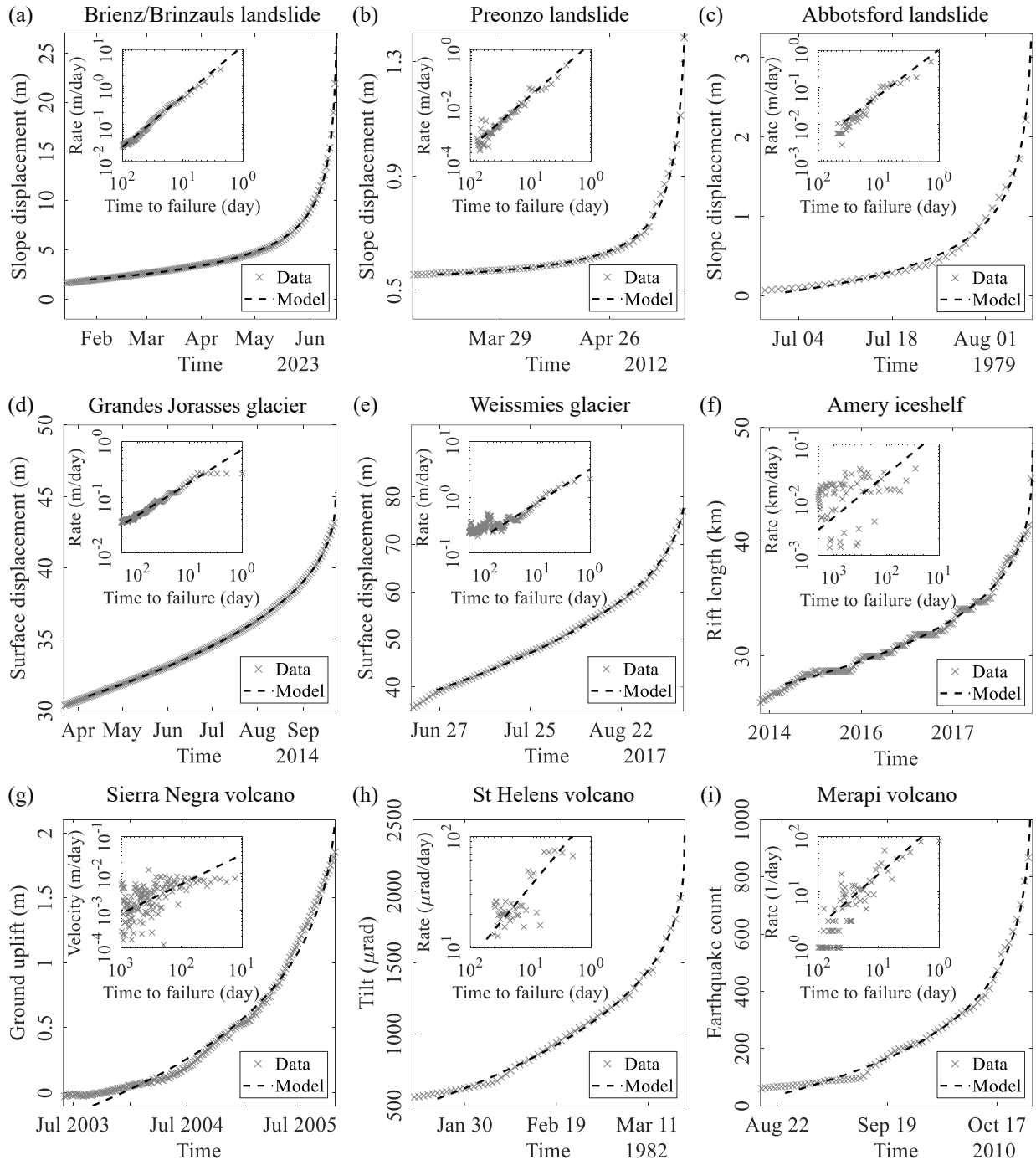

**Figure 1.** Precursory acceleration leading up to catastrophic failure at (a) Brienz/Brinzauls landslide (Switzerland), (b) Preonzo landslide (Switzerland), (c) Abbotsford landslide (New Zealand) (d) Grandes Jorasses glacier (Italy), (e) Weissmies glacier (Switzerland), (f) Amery iceshelf (Antarctica), (g) Sierra Negra volcano (Ecuador), (h) St Helens volcano (USA), and (i) Merapi volcano (Indonesia). The dashed line indicates the fit using the power law finite-time singularity model, with equation (6) for the cumulative quantity and equation (2) for the rate quantity.

## 3 Model derivation and demonstration

### 3.1. An event-based model for the logarithmic finite-time singularity

Let us adopt an event-based perspective and consider the system evolution as a sequence of discrete damage events, such that $\Omega(t)$ accumulates according to:

$$\Omega(t_n) = \Omega(t_{n-1}) + \Delta\Omega_n, \tag{7}$$

where $\Delta\Omega_n$ is the increment associated with the $n$th event, and $n \geq 1$ is an integer indexing the sequence of event occurrences. Iterating equation (7), we obtain:

$$\Omega(t_n) = \Omega(t_0) + \sum_{k=1}^{n} \Delta\Omega_k. \tag{8}$$

We then assume that the event times $t_n$, marking the occurrence of successive increments, follow a pattern of progressively decreasing intervals, that is, events occur more and more frequently as the system evolves. The specific law describing the decreasing intervals between events is taken to be a simple geometrical series with a characteristic scaling ratio $\lambda$:

$$t_{n+1} - t_n = \frac{1}{\lambda}(t_n - t_{n-1}), \text{ with } \lambda > 1. \tag{9}$$

Iterating equation (9) yields:

$$t_n - t_0 = \sum_{k=1}^{n}(t_k - t_{k-1}) = \sum_{k=1}^{n}\frac{t_1 - t_0}{\lambda^{k-1}} = \frac{1 - 1/\lambda^n}{1 - 1/\lambda}(t_1 - t_0). \tag{10}$$

Thus, there is a critical time $t_c$, corresponding to the end of this accelerating sequence of events, which is given by:

$$t_c - t_0 = \lim_{n \to \infty}(t_n - t_0) = \frac{\lambda}{\lambda - 1}(t_1 - t_0). \tag{11}$$

In the limit $n \to \infty$, the interevent time shrinks to zero, indicating that events occur with increasing frequency as the system approaches the critical time $t_c$. This progressive densification of events with shrinking interevent time intervals is a hallmark of a finite-time singularity, defined below. The critical time $t_c$ depends on both the initial interevent time scale $t_1 - t_0$ and the scaling ratio $\lambda$ that controls the time contraction rate. It is important to note that, as $\lambda \to 1$, $t_c$ tends to infinity, as expected.

Dividing equation (11) by equation (10) with some re-arrangements yields:

$$\frac{t_c - t_n}{t_c - t_0} = \frac{1}{\lambda^n}, \tag{12}$$

showing that the succession of times-to-failure $t_c - t_n$ of the sequence of events also form a geometrical time series, contracting to 0 as the $n$th event time $t_n$ converges towards the critical time $t_c$. Rearranging equation (12) gives the expression of $n$ as a function of the time-to-failure $t_c - t_n$:

$$n = \frac{1}{\ln(1/\lambda)} \ln\left(\frac{t_c - t_n}{t_c - t_0}\right) = \log_{1/\lambda}\left(\frac{t_c - t_n}{t_c - t_0}\right). \tag{13}$$

Let us first consider the simplest situation in which each event produces the same observable increment $\Delta\Omega_n$ of the observable quantity $\Omega$, i.e., $\Delta\Omega_n \equiv \delta$. Then, equation (11) reduces to:

$$\Omega(t_n) = \Omega(t_0) + n\delta . \qquad (14)$$

By substituting equation (13) into equation (14) and replacing $t_n$ with the continuous time variable $t$, we obtain:

$$\Omega(t) = \Omega(t_0) + \delta \log_{1/\lambda}\left(\frac{t_c - t}{t_c - t_0}\right). \qquad (15)$$

By further defining $\kappa = \delta/\ln\lambda$, we recover the solution $\Omega(t)$ for $m = 0$ given by equation (4), which is a special case of the finite-time singularity with a logarithmic divergence at $t_c$. This present model thus offers a simple physical picture in which the logarithmic finite-time singularity emerges from the progressive shortening of the waiting time between events of equal magnitude (or comparable magnitudes in an approximate sense). Figure 2 illustrates the temporal evolution of observable $\Omega$ for a range of $\lambda$ values, where the analytical solution from equation (15) provides an accurate description of the numerical simulation results. For larger values of $\lambda$, the critical time $t_c$ of the logarithmic finite-time singularity is earlier, as expected from equation (11).

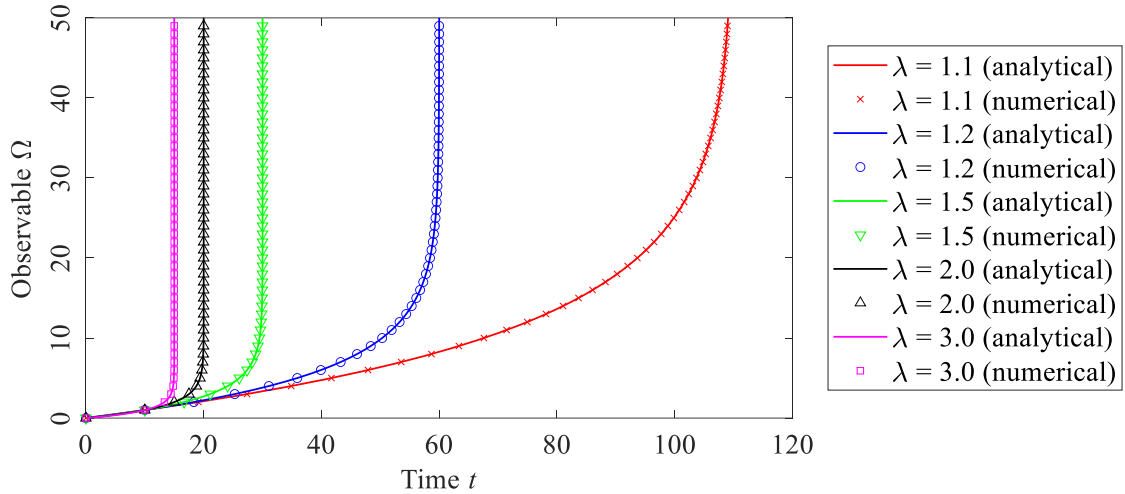

**Figure 2.** Temporal evolution of the dimensionless observable $\Omega$ as a function of dimensionless time $t$, for various time scaling ratios $\lambda$ and a fixed increment $\Delta\Omega$, exhibiting logarithmic finite-time singularities. Model parameters are: initial time $t_0 = 0$, initial observable value $\Omega(t_0) = 0$, first interevent duration $t_1 - t_0 = 10$, increment $\Delta\Omega \equiv \delta = 1$, and time scaling ratios $\lambda = 1.1, 1.2, 1.5, 2.0,$ and $3.0$. Analytical solutions from equation (15) are compared with numerical simulations.

Let us now consider a more general scenario where the increments are independent and identically distributed (i.i.d.) random variables drawn from a certain probability distribution $\mathbb{P}(\Delta\Omega)$. In this stochastic generalization, the event times still follow the deterministic sequence leading to the same logarithmic finite-time singularity as given by equation (11). Consequently, $\Omega(t)$ remains singular at $t_c$, but exhibits fluctuations around the deterministic trajectory. For distributions $\mathbb{P}(\Delta\Omega)$ that have a finite mean, the average trajectory is simply given by equation (15) with $\delta$ replaced by the expectation value of the random increments $\langle\Delta\Omega\rangle$. However, for distributions $\mathbb{P}(\Delta\Omega)$ that have an infinite mean, the dynamics of $\Omega(t)$ is dominated by large fluctuations. In the special case where $\mathbb{P}(\Delta\Omega)$ is a power law:

$$\mathbb{P}(\Delta\Omega) \sim (\Delta\Omega)^{-(1+\mu)}, \text{ with } \mu < 1, \tag{16}$$

the sum of the random variables $\Delta\Omega$ tends to be dominated by large events of size $\sim n^{1/\mu}$. This implies that the cumulative observable also grows anomalously as $\sim n^{1/\mu}$ (Sornette, 2006), such that:

$$\Omega(t_n) \approx \Omega(t_0) + \xi n^{1/\mu}, \tag{17}$$

where $\xi$ is a stochastic variable independent of $n$ and related to the heavy-tailed statistics of the increments. As a result, the observable $\Omega(t)$ exhibits strong fluctuations, with its typical trajectory approximately following:

$$\Omega(t) \approx \Omega(t_0) + \xi \left[ \log_{1/\lambda} \left( \frac{t_c - t}{t_c - t_0} \right) \right]^{1/\mu}, \tag{18}$$

which is obtained by substituting equation (13) into equation (17) and replacing $t_n$ with the continuous time variable $t$.

Figure 3a illustrates the temporal evolution of observable $\Omega$ for Gaussian-distributed increments $\Delta\Omega$, which is well captured by the deterministic trajectory, given by the analytical solution in equation (15), with $\delta$ replaced by the mean of random increments $\langle\Delta\Omega\rangle$. As the standard deviation of $\Delta\Omega$ increases, stronger fluctuations around this deterministic trajectory are observed. Figure 3b displays the temporal evolution of observable $\Omega$ for power law-distributed increments $\Delta\Omega$ with exponents $\mu = 0.9$ and $0.6$. The trajectory $\mu = 0.9$ remains well captured by equation (17), whereas the $\mu = 0.6$ case (with a heavier power law tail) shows noticeable deviations from the analytical solution during the early stage but gradually converges towards it at later times, consistent with the underlying asymptotic assumption. Here, the parameter $\xi$ in equation (17) is calibrated using the Levenberg-Marquardt algorithm based on the numerical simulation data.

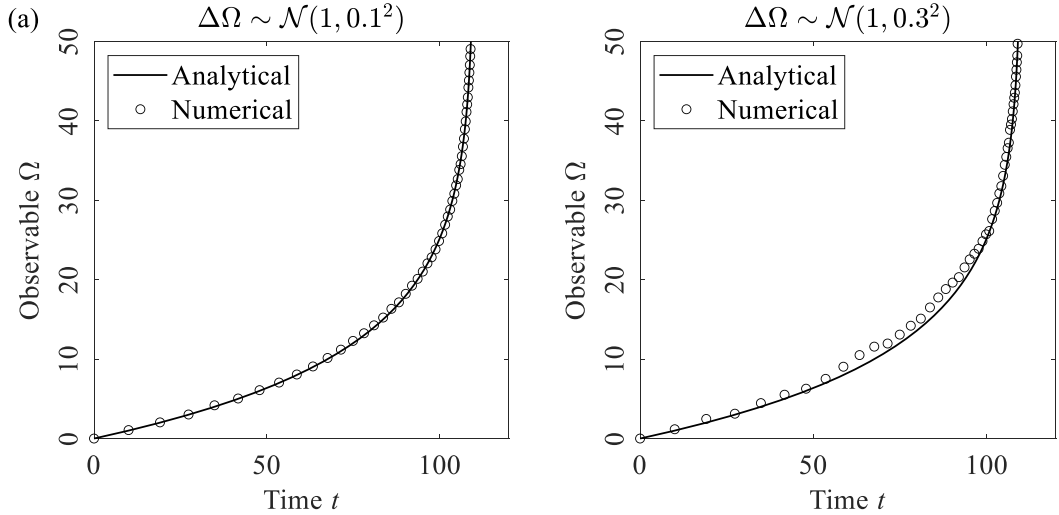

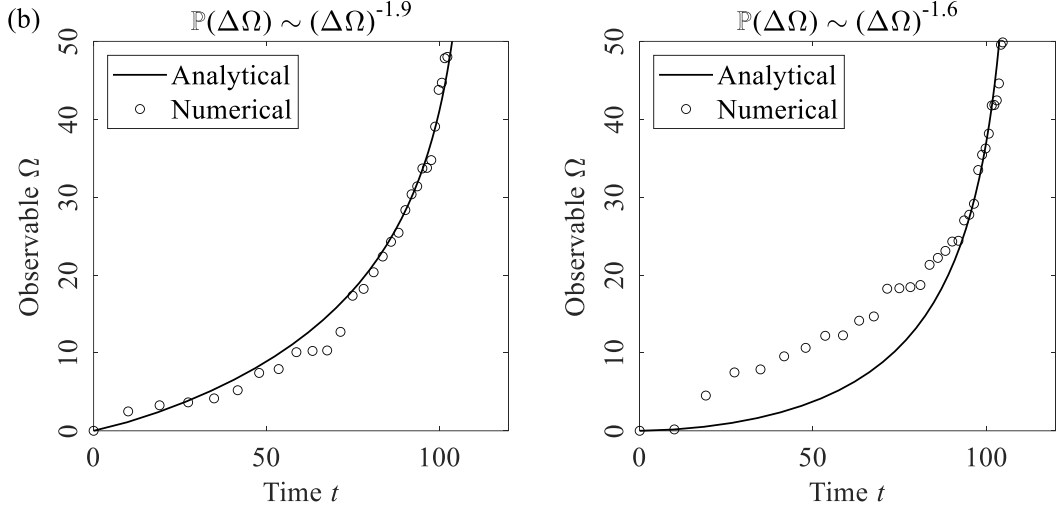

**Figure 3.** Temporal evolution of the dimensionless observable $\Omega$ as a function of the dimensionless time $t$, for (a) Gaussian-distributed and (b) power law-distributed increments $\Delta\Omega$ and a fixed time scaling ratio $\lambda$, exhibiting logarithmic finite-time singularities. Model parameters are: initial time $t_0 = 0$, initial observable value $\Omega(t_0) = 0$, first interevent duration $t_1 - t_0 = 10$, and scaling ratio $\lambda = 1.1$. The increment $\Delta\Omega$ is sampled from either (a) a Gaussian distribution with a mean of 1 and a standard deviation of 0.1 or 0.3, or (b) from a truncated power law distribution with exponent $\mu = 0.9$ or 0.6, with lower and upper bounds of 0.05 and 5, respectively. Analytical solutions from equation (18) are compared with numerical simulations.

3.2. An event-based model for the power law finite-time singularity

Let us extend the model presented in section 3.1 by assuming that the increments also grow (or shrink) geometrically, i.e., $\Delta\Omega_n = \rho^{n-1}\delta$, where $\delta$ is the initial increment and $\rho > 0$ is the geometrical growth factor. Note that, for $\rho > 1$, the increments increase with $n$; for $0 < \rho < 1$, they decrease; and for $\rho = 1$, the increments remain constant. Then, equation (7) becomes:

$$\Omega(t_n) = \Omega(t_{n-1}) + \delta\rho^{n-1}, \qquad (19)$$

where the interevent times remain governed by equation (9). The solution of equation (19) is:

$$\Omega(t_n) = \Omega(t_0) + \delta\sum_{k=1}^{n-1}\rho^k = \Omega(t_0) + \delta\frac{\rho^n - 1}{\rho - 1}. \qquad (20)$$

By substituting equation (13) into equation (20), applying the logarithmic change-of-base formula, and further replacing $t_n$ with the continuous time variable $t$, we obtain:

$$\Omega(t) = \Omega(t_0) + \frac{\delta}{\rho - 1}\left[\left(\frac{t_c - t}{t_c - t_0}\right)^{\log_{1/\lambda}\rho} - 1\right], \qquad (21)$$

which recovers the formulation of the power law finite-time singularity, i.e., equation (6), by defining:

$$A = \Omega(t_0) - \frac{\delta}{\rho - 1}, \qquad (22)$$

$$B = \frac{\delta}{(\rho-1)(t_c - t_0)^m}, \tag{23}$$

$$m = \log_{1/\lambda} \rho = -\frac{\ln \rho}{\ln \lambda}. \tag{24}$$

The critical exponent $m$ encapsulates the interplay between the geometrical contraction of interevent times (characterized by $\lambda$) and the geometrical growth of event magnitudes (characterized by $\rho$). Note that as $\rho \to 1$, the exponent $m \to 0$, thereby recovering the logarithmic singularity described by equation (15). Given $\lambda > 1$, we observe that $m < 0$ when $\rho > 1$, where the increments are increasing over time; and $0 < m < 1$ when $\lambda^{-1} < \rho < 1$, where the increments are decreasing over time. Thus, a power law finite-time singularity is expected when $\rho > \lambda^{-1}$, below which no divergence of $\Omega$ and $d\Omega/dt$ occur at $t_c$ (however, higher-order derivatives of $\Omega$ can still diverge).

This provides the following intuitive physical understanding. First, for $\Omega$ and/or $d\Omega/dt$ to exhibit a finite-time singularity, the frequency of events must accelerate so that there is an infinite number of events in finite time with a convergence of the event occurrence times at the critical time $t_c$. The second differentiating property is how the increments of the observable behave on the approach to the critical time $t_c$:

- When the increments decay according to a geometric series with scaling ratio $\rho$ (with $\lambda^{-1} < \rho < 1$), $\Omega$ exhibits a finite-time singularity at $t_c$ due to an infinite slope at $t_c$ (i.e., $d\Omega/dt$ diverges), while $\Omega$ converges to a finite value $A$ at $t_c$.
- When the increments are constant ($\rho = 1$), $\Omega$ exhibits a logarithmic finite-time singularity at $t_c$.
- When the increments grow according to a geometric series with scaling ratio $\rho > 1$, $\Omega$ exhibits a finite-time singularity with a power law divergence at $t_c$.

Equation (21) can also be extended to a more general scenario where the geometrical growth factor $\rho$ is drawn from a probability distribution $\mathbb{P}(\rho)$, while we still assume that the event times follow the deterministic sequence given by equation (11). Then, $\Delta\Omega_n = \rho^{n-1}\delta$ is replaced by $\Delta\Omega(t_n) = \rho_{n-1}\Delta\Omega(t_{n-1}) = \delta\rho_1\rho_2\ldots\rho_{n-1}$, which assumes that successive increments are governed by a simple multiplicative relationship, reflecting a strong memory effect between consecutive damage or deformation events. In this framework, each microcracking event or deformation increment is conditioned by the preceding one, suggesting that the subsequent increment is both initiated and scaled by its predecessor. Equation (19) then becomes $\Omega(t_n) = \Omega(t_{n-1}) + \Delta\Omega(t_n) = \Omega(t_{n-1}) + \delta\rho_1\rho_2\ldots\rho_{n-1}$. Introducing the change of notation $r_j = \rho_{n-j}$, we obtain: $\Omega(t_n) = \delta(1 + r_{n-1} + r_{n-1}r_{n-2} + r_{n-1}r_{n-2}\ldots r_1)$, so that $\Omega(t_n) = \delta + r_{n-1}\Omega(t_{n-1})$. This stochastic recurrence equation represents a special case of the Kesten multiplicative problem (Buraczewski et al., 2016; Calan et al., 1985; Kesten, 1973; Sornette, 2006), which gives rise to remarkably rich behaviors (Sornette, 1998).

For $\langle \ln\rho \rangle = \langle \ln r \rangle < 0$, meaning that the random scaling ratios $r_n$'s are typically smaller than 1 but not necessarily always, $\Omega(t_n)$ has the structure of a stable auto-regressive process with stochastic regression coefficient $r_n$. For $n \to \infty$, $\Omega(t_n)$ converges to a random variable $\Omega_\infty$ with a power law distribution $\mathbb{P}(\Omega) \sim 1/\Omega^{1+\nu}$, for large $\Omega$, where the exponent $n$ is the solution of $\langle r^\nu \rangle = 1$ (Kesten, 1973; Sornette & Cont, 1997). This distribution describes the ensemble of possible final values of the observable $\Omega$ at their critical time over the set of possible realizations of the sequences of random multiplicative terms $r_1, r_2, \ldots, r_n, \ldots$, i.e., over all possible histories of the damage or deformation process. The specific temporal evolution of the stochastic $\Omega(t_n)$ as it approaches its terminal value $\Omega_\infty$ at the critical time $t = t_c$ can be obtained by noting that $\Omega_\infty - \Omega(t_n) = \delta(r_n r_{n-1} \ldots r_1 + r_{n+1} r_n r_{n-1} \ldots r_1 + \ldots) = \delta r_n r_{n-1} \ldots r_1 (1 + r_{n+1} + r_{n+1} r_{n+2} + \ldots)$, which can be written as $\Omega_\infty - \Omega(t_n) = r_n r_{n-1} \ldots r_1 \Omega_\infty^{(n+1)}$ with $\Omega_\infty^{(n+1)} = \delta(1 + r_{n+1} + r_{n+1} r_{n+2} + \ldots)$ having the same form as $\Omega_\infty = \lim_{n \to \infty} \Omega(t_n)$ (with an infinite number of products) but with $r_1$ replaced by $r_{n+1}$, $r_2$ replaced by $r_{n+2}$, and so on. Since $\langle \ln \rho \rangle = \langle \ln r \rangle < 0$, we can write $r_n = \exp[n \langle \ln r_n \rangle + \varsigma \sqrt{n} + O(1)]$, where $\varsigma$ is a random variable of order 1. To leading order, using equation (13), we obtain $\Omega_\infty - \Omega(t) = \Omega_\infty^{(n+1)} \left[ (t_c - t_n) / (t_c - t_0) \right]^{-\langle \ln \rho \rangle / \ln \lambda}$, which recovers equation (21) for $\lambda^{-1} < \rho < 1$, except for two distinct stochastic features: (i) the prefactor $\Omega_\infty^{(n+1)}$ of the power law term corresponds to a random amplitude fixed for any individual realization of the system; (ii) the asymptotic value $\Omega_\infty$ of the observable at the critical time is also a random variable but fixed for any individual realization of the system.

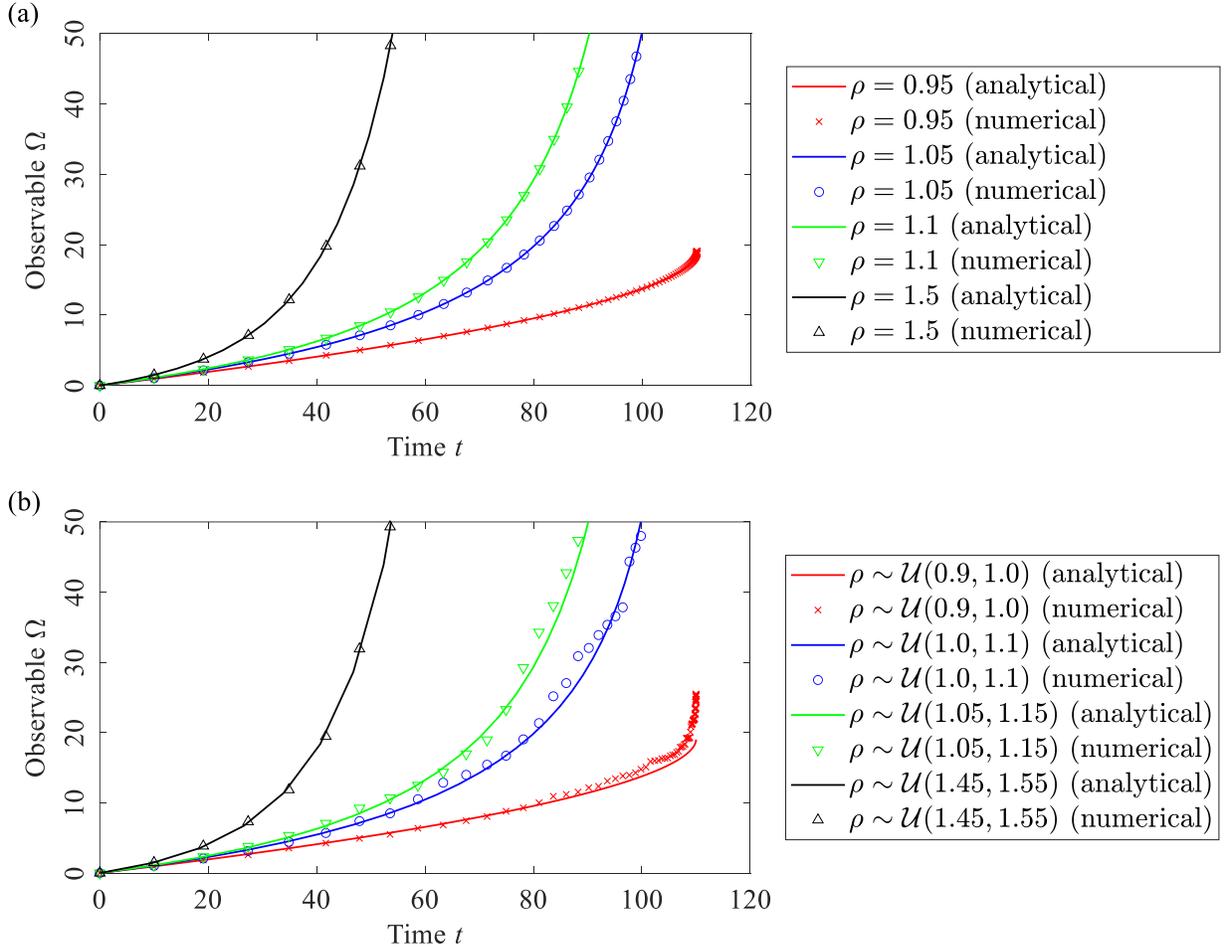

**Figure 4.** Temporal evolution of dimensionless observable $\Omega$ as a function of dimensionless time $t$, for geometrically growing increments $\Delta\Omega$ and a fixed time scaling ratio $\lambda$, exhibiting power law finite-time singularities. Model parameters are: initial time $t_0 = 0$, initial observable value $\Omega(t_0) = 0$, first interevent duration $t_1 - t_0 = 10$, time scaling ratio $\lambda = 1.1$, and initial increment $\delta = 1$. Two scenarios are considered for the geometrical growth factor $\rho$: (a) it takes fixed values of 0.95, 1.05, 1.1, or 1.5 or (b) it follows a uniform distribution with a mean of 0.95, 1.05, 1.1, or 1.5 and an interval size of 0.1. Analytical solutions from equation (21) are compared with numerical simulations.

The other case of interest occurs for $\langle \ln\rho\rangle = \langle \ln r\rangle > 0$, meaning that the random scaling ratios $r_n$'s are typically larger than 1, leading to a (stochastic) divergence of the observable. Coming back to equation $\Omega(t_n) = \delta + r_{n-1}\Omega(t_{n-1})$, as $\Omega(t_n)$ grows very fast without bounds on the approach to $t_c$, the term $\delta$ becomes negligible and $\Omega(t_n)$ reduces to a very good approximation to the product $\Omega(t_n) = \delta r_{n-1}...r_1 = \delta \exp[n\langle \ln r\rangle + \varsigma\sqrt{n} + O(1)]$. Using equation (13), we obtain the leading term describing the finite time singularity regime in which $\Omega(t_n)$ diverges as $t$ tends to $t_c$ according to $\Omega(t) = \delta\left[(t_c - t_n)/(t_c - t_0)\right]^{-\langle \ln\rho\rangle/\ln\lambda}$, which also recovers equation (21) for $\rho > 1$.

Figure 4 displays the temporal evolution of the observable $\Omega$ resulting from geometrically growing increments $\Delta\Omega$, where the growth factor $\rho$ either takes fixed values (Figure 4a) or is sampled from a uniform distribution (Figure 4b). In the deterministic case (Figure 4a), the analytical solution given by equation (21) closely matches the numerical results. As predicted, $\Omega$ diverges at $t_c$ for cases with $\rho > 1$, but for $\rho < 1$ converges at $t_c$ to the finite value $A$ given by equation (22). It is worth reminding that $t_c$ is nothing but the time around which the system transitions into the elasto-dynamical regime, where the rupture velocity becomes very large and limited only by the speed of elastic waves. In the stochastic case (Figure 4b), the analytical solution still well captures the overall trend, despite the presence of fluctuations due to randomness in $\rho$.

### 3.3. A generalized stochastic event-based model of finite-time singularities

We further generalize the deterministic recurrence relation, equation (9), into a stochastic one, by allowing the time scaling ratio $\lambda$ to be an i.i.d. random variable:

$$t_{n+1} - t_n = \frac{1}{\lambda_n}(t_n - t_{n-1}), \text{ with } \lambda_n > 1. \qquad (25)$$

This expression (25) accounts for fluctuations in the acceleration of event rates, providing a minimal yet effective model to capture the intrinsic stochasticity of damage processes in heterogeneous materials with complex rheology. Iterating equation (25), we obtain:

$$t_n - t_0 = (t_1 - t_0)\sum_{k=0}^{n-1}\left(\prod_{j=1}^{k}\frac{1}{\lambda_j}\right) = \zeta_n(t_1 - t_0), \qquad (26)$$

with the empty product being defined as unity. The sum of product terms defines the stochastic variable:

$$\zeta_n = 1 + \frac{1}{\lambda_1} + \frac{1}{\lambda_1\lambda_2} + ... + \frac{1}{\lambda_1\lambda_2...\lambda_{n-1}}, \qquad (27)$$

which enjoys a nested hierarchical structure that can be seen by reparametrizing $\chi_{n-j}=1/\lambda_j$ leading to the recurrence relation $\zeta_{n+1} = 1 + \chi_n\zeta_n$. We recognize again a stochastic recurrence equation of the form of the Kesten multiplicative problem (Buraczewski et al., 2016; Calan et al., 1985; Kesten, 1973; Sornette, 2006), which is characterized by rich and intricate behaviors (Sornette, 1998). For $\langle\ln(\chi_n)\rangle < 0$ (which corresponds to $\langle\ln(\lambda_n)\rangle > 0$, meaning that the random scaling ratios $\lambda_n$'s are typically larger than 1, but not necessarily always), $\zeta_n$ has the structure of a stable auto-regressive process with stochastic regression coefficient $\chi_n$. For $n\to\infty$, $\zeta_n$ converges to a random variable $\zeta$ with a power law distribution:

$$\mathbb{P}(\zeta) \sim 1/(\zeta)^{1+\upsilon}, \text{ for large } \zeta, \tag{28}$$

where the exponent $\upsilon$ is the solution of (Kesten, 1973; Sornette & Cont, 1997):

$$\langle \chi^\upsilon \rangle = \langle (1/\lambda)^\upsilon \rangle = 1. \tag{29}$$

The critical time $t_c$ at which a finite-time singularity occurs can be then obtained from:

$$t_c - t_0 = \lim_{n\to\infty}(t_n - t_0) = \zeta_\infty (t_1 - t_0), \tag{30}$$

and is thus a stochastic variable changing from realization to realization. Given the condition $\langle \ln(1/\lambda_n) \rangle < 0$ mentioned above and the i.i.d. nature of the random variables $\lambda_n$'s, the average critical time can be calculated as:

$$\langle t_c \rangle - t_0 = \frac{1}{1 - \langle 1/\lambda \rangle}(t_1 - t_0). \tag{31}$$

which generalizes expression (11). Since $t_c - t_0$ is proportional to $\zeta_\infty$ according to expression (30), the probability distribution given by (28) with (29) also describes the distribution of critical times. When $\lambda$ (and consequently $1/\lambda$) is lognormally distributed, equation (29) for $\upsilon$ can be analytically solved via a Gaussian integral (Lei & Sornette, 2023; Sornette & Cont, 1997), giving:

$$\upsilon = -\frac{2\langle \ln(1/\lambda) \rangle}{\text{var}[\ln(1/\lambda)]} = \frac{2\langle \ln \lambda \rangle}{\text{var}(\ln \lambda)}. \tag{32}$$

If $\lambda$ is not lognormally distributed, the above equation only gives an approximation for $\upsilon$, while a more precise solution can be obtained by numerically solving equation (29). Equation (31) holds only when the expectation value of $t_c$ is finite, which requires $\upsilon > 1$. For $\upsilon < 1$, the distribution of critical times is heavy-tailed, so that the notion of an average critical time to rupture breaks down. On the other hand, a median critical time and its various quantiles continue to be well-defined and can be obtained from numerical simulations of the Kesten process.

Combining equations (26) and (30), we obtain:

$$\frac{t_c - t_n}{t_c - t_0} = \frac{\zeta_\infty - \zeta_n}{\zeta_\infty}, \tag{33}$$

with

$$\zeta_\infty - \zeta_n = \zeta_\infty^{n+1} \prod_{i=1}^n \chi_i, \tag{34}$$

where we denote

$$\zeta_\infty^{n+1} = 1 + \chi_{n+1} + \chi_{n+1}\chi_{n+2} + \cdots + \chi_{n+1}\chi_{n+2}\cdots\chi_{n+j} + \cdots, \tag{35}$$

which is a random variable with the same structure as $\zeta_\infty$ (with an infinite number of products) but with $\chi_1$ replaced by $\chi_{n+1}$, $\chi_2$ replaced by $\chi_{n+2}$, and so on. Combining equations (33) and (34) yields:

$$\frac{t_c - t_n}{t_c - t_0} = \frac{\zeta_\infty^{n+1}}{\zeta_\infty^1} \prod_{i=1}^{n} \chi_i = \frac{1}{\gamma_n} \prod_{i=1}^{n} \chi_i, \qquad (36)$$

where we use the notation $\zeta_\infty^1$ for the limit $n \to \infty$ of the random variable defined by equation (27) to clarify the comparison between $\zeta_\infty^1$ and $\zeta_\infty^{n+1}$. Here, for conciseness, we further denote the ratio of these two heavy-tailed random variables as $\gamma_n$. This random variable also follows a power law distribution with the same exponent $\upsilon$ as determined by the solution of equation (29). This arises because $1/\zeta_\infty^1$ has a thin-tailed distribution and, for large $n$, the dominant contributions to large excursions of $\zeta_\infty^{n+1}$ stem from products in the numerator that become asymptotically independent from the products in the sum forming the denominator. Using the Central Limit Theorem, it is easy to show that the two leading orders in powers of $\sqrt{n}$ for $\ln\left(\prod_{i=1}^{n} \chi_i\right)$ yield:

$$\prod_{i=1}^{n} \chi_i \approx \exp\left[n\langle \ln(1/\lambda) \rangle + \sigma\sqrt{n} Z\right], \qquad (37)$$

where $\sigma$ is the standard deviation of the random variable $\ln\lambda$ and $Z$ is a standard random variable (normally distributed with zero mean and unit variance). Combining equations (36) and (37), and keeping only the leading order gives:

$$n \approx \frac{1}{\langle \ln(1/\lambda) \rangle} \ln\left(\gamma_n \frac{t_c - t_n}{t_c - t_0}\right), \qquad (38)$$

Inserting equation (38) into equation (14), and replacing $t_n$ and $\gamma_n$ by the continuous variables $t$ and $\gamma$, we obtain:

$$\Omega(t) \approx \Omega(t_0) + \frac{\delta}{\langle \ln(1/\lambda) \rangle} \ln\left(\gamma \frac{t_c - t}{t_c - t_0}\right) = \Omega'(t_0) + \frac{\delta}{\langle \ln(1/\lambda) \rangle} \ln\left(\frac{t_c - t}{t_c - t_0}\right), \qquad (39)$$

with

$$\Omega'(t_0) = \Omega(t_0) + \frac{\delta}{\langle \ln(1/\lambda) \rangle} \ln \gamma. \qquad (40)$$

Comparing equation (39) with equation (15), the fact that the scaling ratios $\lambda_n$'s are random leads to (i) replacing $\ln\lambda$ by $\langle \ln\lambda \rangle$, (ii) introducing a stochastic critical time $t_c$ according to equation (30), and (iii) using a stochastic reference observable $\Omega'(t_0)$. These equations can be further generalized to account for a stochastic geometrical growth factor $\rho$ drawn from a probability distribution $\mathbb{P}(\rho)$.

Similar equations can be obtained by inserting (38) into equations (17) and (20). Then, instead of having equation (17), we obtain:

$$\Omega(t) \approx \Omega(t_0) + \frac{\xi}{\langle \ln(1/\lambda) \rangle^{1/\mu}} \left[\ln\left(\gamma \frac{t_c - t}{t_c - t_0}\right)\right]^{1/\mu}, \qquad (41)$$

Instead of equation (21), we have:

$$\Omega(t) \approx \Omega(t_0) + \frac{\delta}{\rho - 1}\left[\left(\gamma \frac{t_c - t}{t_c - t_0}\right)^{m'} - 1\right]. \qquad (42)$$

with the critical exponent $m'$ given by:

$$m' = -\frac{\ln \rho}{\langle \ln \lambda \rangle}. \tag{43}$$

Compared to their deterministic counterparts, note that equations (39)-(42) incorporate the random variable $\gamma$. As shown in equation (40), this introduces a stochastic component in the final value of the observable at $t_c$. Furthermore, equation (42) demonstrates that $\gamma$ gives rise to a stochastic prefactor governing the scaling of the power law singularity.

Figure 5a illustrates the temporal evolution of observable $\Omega$ for lognormally distributed time scaling ratios $\lambda$ with a fixed increment $\Delta\Omega \equiv \delta = 1$. A total of 50 random realizations is simulated numerically, and their ensemble mean is well captured by the analytical solution given by equation (39). As the variance of $\ln\lambda$ increases, variability between different realizations becomes more pronounced, highlighting that the randomness in $\lambda$ has a strong influence on the critical time $t_c$ associated with individual trajectories. We further analyze the temporal evolution of observable $\Omega$ for lognormally distributed time scaling ratios $\lambda$, while the increments $\Delta\Omega$ follow either a Gaussian distribution (Figure 5b) or a power law distribution (Figure 5c). For Gaussian-distributed $\Delta\Omega$, the mean trend of $\Omega$ is well captured by the typical solution given by equation (41), despite the strong variability across different realizations. For power law-distributed $\Delta\Omega$, the $\Omega$ trajectories exhibit more pronounced temporal fluctuations and greater variability across different realizations, compared to those of fixed or Gaussian distributed $\Delta\Omega$. The ensemble averaged $\Omega$ trajectory obtained from numerical simulations remains well described by equation (41) for $\mu = 0.9$, whereas for $\mu = 0.6$, it shows considerable deviations from the analytical solution in the early stage but gradually converges towards it over time, in line with the underlying asymptotic assumption. Here, the parameter $\xi/\langle\ln(1/\lambda)\rangle^{1/\mu}$ in equation (41) is calibrated against the ensemble averaged $\Omega$ trajectory using the Levenberg-Marquardt algorithm. Figure 6 examines the analytical solution given by equation (42) for describing the temporal evolution of observable $\Omega$, assuming lognormally distributed time scaling ratios $\lambda$ and geometrically grown increments $\Delta\Omega$. The analytical prediction agrees well with the ensemble average of the numerical simulations, thereby validating equation (42). The interplay between the randomness in the time scaling ratio $\lambda$ and the randomness in the geometrical growth factor $\rho$ appears to give rise to an oscillatory behavior in the $\Omega$ trajectories.

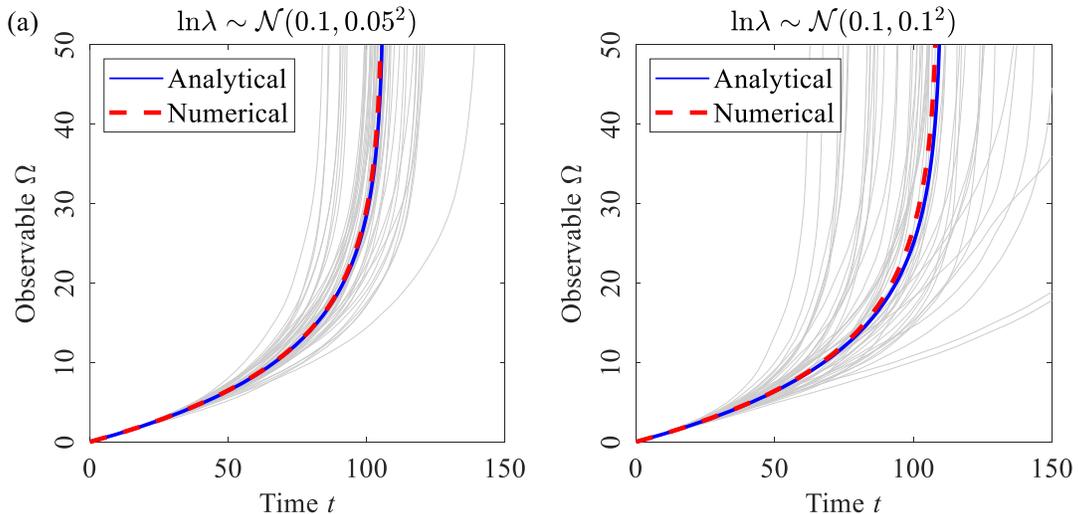

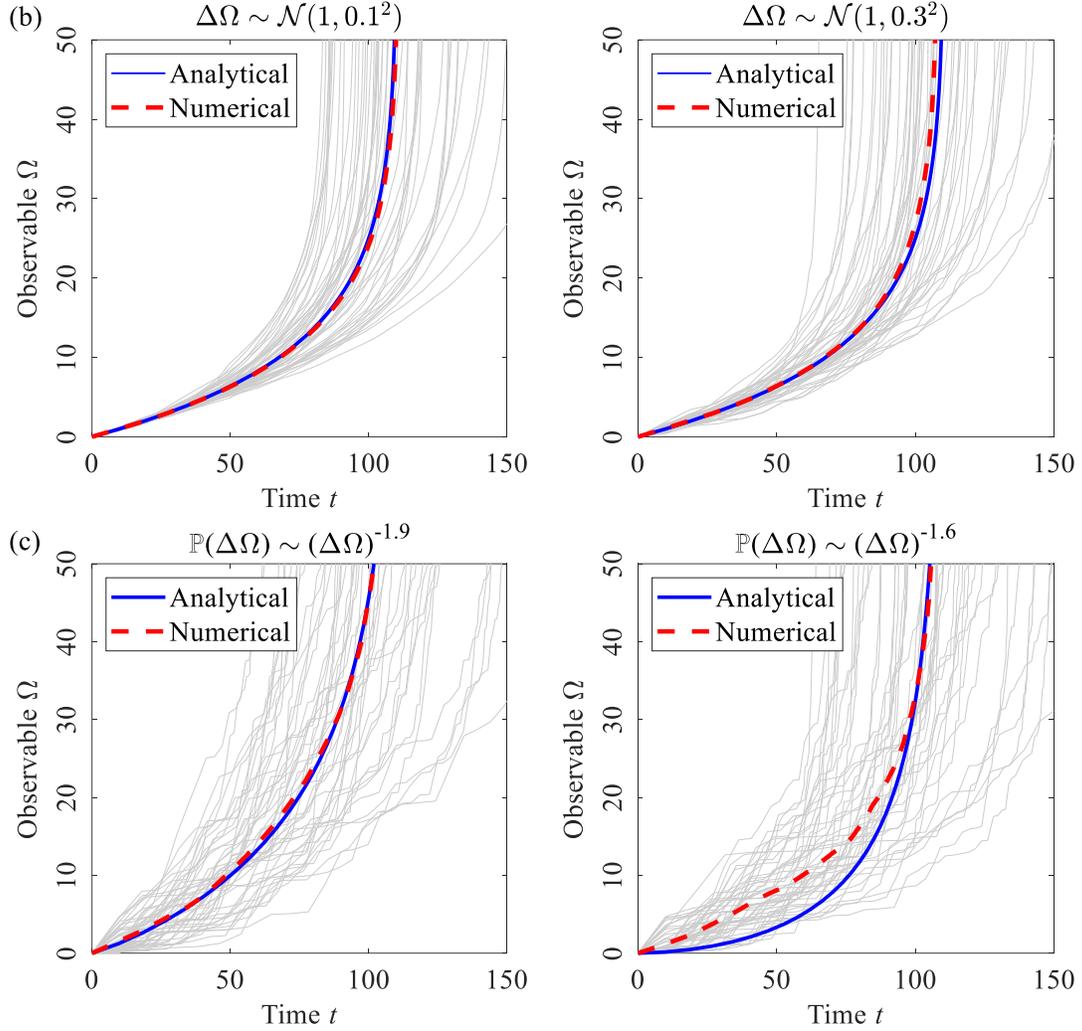

**Figure 5.** Temporal evolution of the dimensionless observable Ω as a function of dimensionless time $t$, for stochastic time contraction ratios $\lambda$, exhibiting logarithmic finite-time singularities. Model parameters are: initial time $t_0 = 0$, initial observable value $\Omega(t_0) = 0$, and first interevent duration $t_1 - t_0 = 10$. The increment $\Delta\Omega$ either (a) takes a fixed value of 1, or is drawn (b) from a Gaussian distribution with a mean of 1 and a standard deviation of 0.1 or 0.3, or (c) from a truncated power law distribution with exponent $\mu = 0.9$ or $0.6$, and lower and upper bounds of 0.05 and 5, respectively. The log time contraction ratio $\ln\lambda$ follows a Gaussian distribution, with a mean of 0.1 and a standard deviation of 0.05 or 0.1 in (a), and 0.1 in both (b) and (c). Analytical solutions from equation (39) are compared with numerical simulations, with grey lines displaying 50 random realizations and the dashed line representing their ensemble mean.

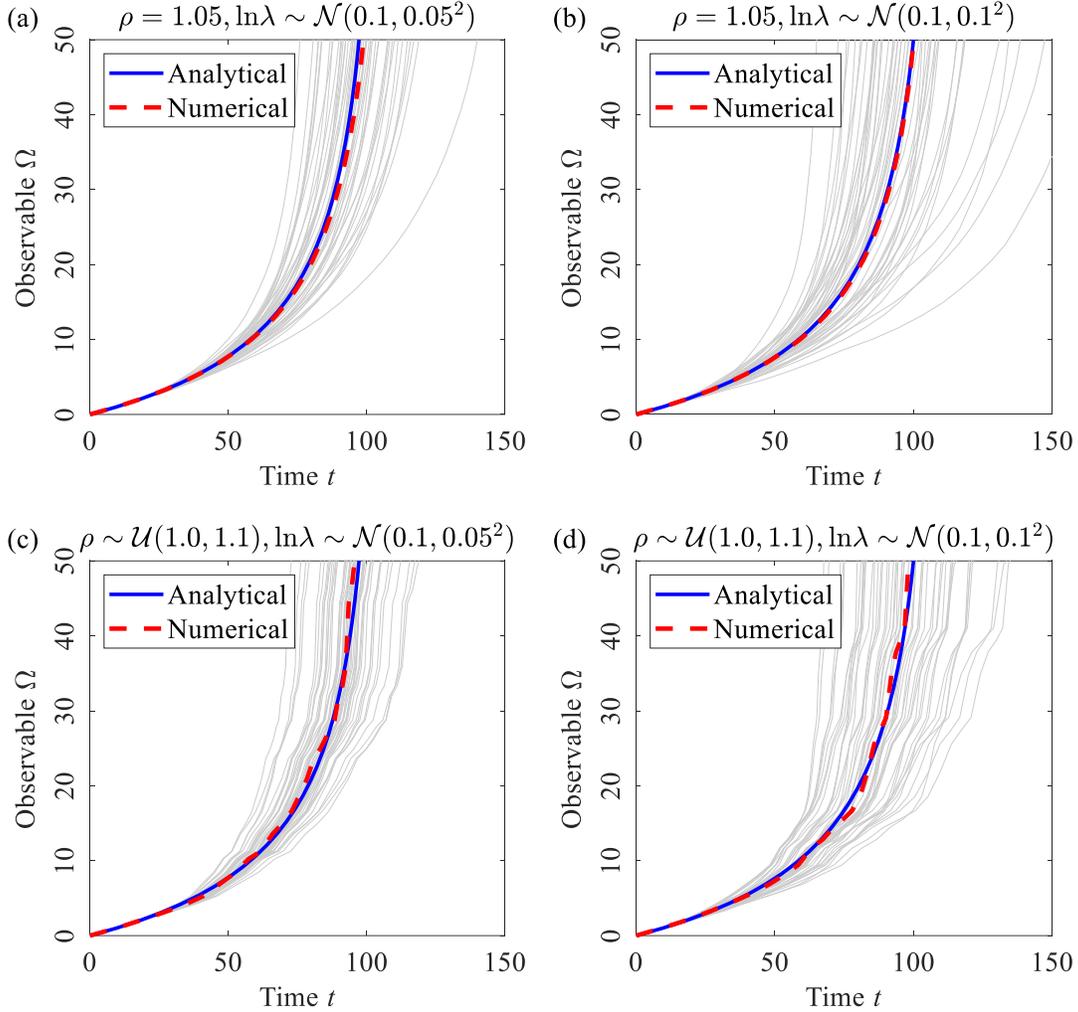

**Figure 6.** Temporal evolution of dimensionless observable Ω as a function of dimensionless time $t$, for stochastic time contraction ratios $\lambda$ and geometrically grown increments $\Delta\Omega$, exhibiting power law finite-time singularities. Model parameters are: initial time $t_0 = 0$, initial observable value $\Omega(t_0) = 0$, first interevent duration $t_1 - t_0 = 10$, and initial increment $\delta = 1$. The geometrical growth factor $\rho$ takes a fixed value of 1.05 in (a) and (b), or follows a uniform distribution with a mean of 1.05 and an interval size of 0.1 in (c) and (d). The log time scaling ratio $\ln\lambda$ follows a Gaussian distribution, with a mean of 0.1 and a standard deviation of 0.05 in (a) and (c), and 0.1 in (b) and (d). Analytical solutions from equation (41) are compared with numerical simulations, with grey lines displaying 50 random realizations and the dashed line representing their ensemble mean.

## 4 Discussion and conclusion

In this paper, we have developed a suite of reduced event-based models to characterize the temporal evolution of geophysical systems during the terminal acceleration phase preceding catastrophic failure. This perspective provides an intuitive physical interpretation of continuous geophysical time series by focusing on the sequence of discrete damage events that punctuate the system's progression. This approach, akin to the quantum mechanical view of light as discrete quanta rather than a continuous wave, offers novel insights into underlying mechanisms that may remain obscured in purely continuous formulations. By representing the evolution of event timing and magnitude and their impact on observables, the event-based framework explains simply the emergence of distinct classes of finite-time singularities. A logarithmic singularity arises when events recur at an accelerating rate, with interevent times contracting geometrically while event magnitudes remain approximately constant. In contrast, a power law finite-time singularity results from the interplay between geometrically contracting interevent times and multiplicatively evolving event magnitudes. Moreover, when the event amplitudes grow exponentially from one event to the next, the observable diverges at the critical time. When the event amplitudes decrease, the observable does not diverge and it is the first-order derivative (or higher-orders) that diverges at the critical time. Within this framework, we further generalized the formulation to a stochastic setting, enabling analysis of how deterministic event trends interact with intrinsic system randomness. This extension captured the role of stochastic fluctuations in modulating the approach to failure and shaping the emergent singular behavior. These models provide a unifying explanation for a range of observed precursory behaviors and scaling laws in geophysical systems approaching catastrophic failure.

Our theoretical perspective finds support in a wide range of empirical observations spanning laboratory experiments to field-scale measurements. For example, time series of various geophysical observables, such as strain, seismic event count, and cumulative energy release, can often be viewed as aggregates of discrete events. Prior theoretical, experimental, and observational studies have demonstrated that the corresponding rate quantities, i.e. strain rate, event rate, and energy release rate, all exhibit power law acceleration dynamics on the approach to catastrophic failure (Amitrano & Helmstetter, 2006; Bell et al., 2011; Bufe & Varnes, 1993; Jaumé & Sykes, 1999; Nechad et al., 2005a; Vu et al., 2019), consistent with the finite-time singularity framework described above. Their power law time-to-failure scaling behavior can all be described by equation (2), expressed separately as follows:

$$\dot{\varepsilon}(t) \sim \frac{1}{(t_c - t)^{p_\varepsilon}}, \quad \dot{N}(t) \sim \frac{1}{(t_c - t)^{p_N}}, \quad \dot{W}(t) \sim \frac{1}{(t_c - t)^{p_W}}, \qquad (44)$$

where the exponents typically satisfy $p_\varepsilon \approx p_N \approx p_W \approx 1$ 1 (Voight, 1989, 1988; Nechad et al., 2005b; Bell et al., 2011; Saichev & Sornette, 2005). However, deviations from unity have also been frequently observed (Amitrano & Helmstetter, 2006; Bell et al., 2013; Vu et al., 2019; Patton et al., 2023; Lei & Sornette, 2025d; Nechad et al., 2005a). For example, numerical and experimental studies of uniaxial compression on heterogeneous samples have reported $p_\varepsilon \approx p_W \approx 1.3$ and $p_N \approx 0.6\text{-}0.8$ (Amitrano & Helmstetter, 2006; Vu et al., 2019). Below, we introduce a simple mean-field model to explain the relationships among these exponents.

Assume the system has an initial modulus of $E_0$ and consists of $M$ number of equally sized representative elements. Under constant applied stress $\sigma$, the system's modulus decreases as elements progressively fail. The effective modulus $E_n$ after the $n$th damage event is given by:

$$E_n = \frac{E_0}{M}(M - N_n), \qquad (45)$$

where $N_n$ is the cumulative number of damage events. Thus, the total strain after the $n$th event is:

$$\varepsilon_n = \frac{\sigma}{E_n} = \frac{\sigma M}{E_0(M - N_n)}. \qquad (46)$$

Then, the strain rate for the $n$th event is given by:

$$\dot{\varepsilon}_n = \frac{\varepsilon_n - \varepsilon_{n-1}}{t_n - t_{n-1}} = \frac{\sigma M}{E_0(t_n - t_{n-1})} \left( \frac{1}{M - N_n} - \frac{1}{M - N_{n-1}} \right) = \frac{\sigma M}{E_0} \frac{\dot{N}_n}{(M - N_n)(M - N_{n-1})}, \quad (47)$$

denoting $\dot{N}_n = N_n - N_{n-1}$. Assuming damage increments are small, i.e. $N_n \approx N_{n-1}$, substituting equation (46) into equation (47) and changing $t_n$ to the continuous time variable $t$, we obtain:

$$\dot{\varepsilon}(t) \approx \frac{E_0}{\sigma M} \varepsilon(t)^2 \dot{N}(t). \quad (48)$$

From equation (6) and considering the initial condition of $\varepsilon(t = 0) = 0$, we can write:

$$\varepsilon(t) = A \left[ 1 - \text{sign}(m_\varepsilon) \left( \frac{t_c - t}{t_c} \right)^{m_\varepsilon} \right], \quad (49)$$

which can be substituted into equation (48) to yield:

$$\dot{\varepsilon}(t) \approx \frac{E_0 A^2}{\sigma M} \left[ 1 - \text{sign}(m_\varepsilon) \left( \frac{t_c - t}{t_c} \right)^{m_\varepsilon} \right]^2 \dot{N}(t). \quad (50)$$

Let us consider the regime where $t$ approaches $t_c$. Considering equation (44) and the relation of $m_\varepsilon = 1 - p_\varepsilon$, counting the powers of $t_c - t$, we obtain $p_\varepsilon + p_N \approx 2$ for $p_\varepsilon > 1$ ($m_\varepsilon < 0$), and $p_\varepsilon \approx p_N$ for $p_\varepsilon < 1$ ($m_\varepsilon > 0$). This is consistent with the commonly reported values $p_\varepsilon \approx p_N \approx 1$ as well as with the cases where $p_\varepsilon \approx 1.3$ and $p_N \approx 0.6$-$0.8$. The observation that $p_\varepsilon \approx p_W$ can be readily explained by the relationship between energy release and strain, since the incremental energy release is proportional to the product of stress and incremental strain, i.e. $\Delta W = \sigma \Delta \varepsilon / 2$, implying a linear scaling between strain rate and energy release rate (Amitrano & Helmstetter, 2006).

Finally, the relationships among the different power law exponents can be expressed in a general form as:

$$p_\varepsilon = p_W \approx 1 + \theta \quad \text{and} \quad p_N \approx 1 - |\theta|, \quad (51)$$

with $-1 < \theta < 1$. The bounds on $\theta$ ensure that the power law exponents remain positive. It is important to note that these relationships are derived under the assumption of constant stress; deviations may be expected when the applied stress varies with time. Interestingly, for $p_\varepsilon < 1$ ($\theta > 0$), it can be shown that the effective modulus of the system scales as $E \sim (t_c - t)^\theta$ when $t$ is close to $t_c$, with the exponent $\theta$ characterizing the macroscopic modulus weakening of the system as failure is approached. This scaling behavior aligns with previous studies (Andersen et al., 1997; Sornette & Andersen, 1998; Turcotte et al., 2003) and is relevant to systems where failure is driven by global stiffness reduction. Conversely, for $p_\varepsilon > 1$ ($\theta < 0$), the effective modulus of the system would converge to a finite value at $t_c$, indicating that the system undergoes accelerating deformation without dramatic softening of its effective stiffness. This scenario may apply to systems where failure is controlled more by accumulated energy or local stress concentration than by damage accumulation leading to overall modulus degradation. Therefore, our simple mean-field model provides a coherent physical explanation for the observed diversity in power law exponents describing accelerating strain rate, event rate, and energy release rate, by linking them to the evolution of system stiffness prior to failure.

The above derivation can be further extended to consider varying stress,

$$\sigma(t) = \sigma_0 + \beta t, \tag{52}$$

where $\sigma_0$ is the initial stress at time $t = 0$, and $\beta$ is the stress increase rate. Then, the strain is given by:

$$\varepsilon(t) = \frac{\sigma(t)}{E(t)} = \frac{\sigma_0 + \beta t}{E_0} \frac{M}{M - N(t)}, \tag{53}$$

with the strain rate further derived as:

$$\dot\varepsilon(t) = \frac{d\varepsilon(t)}{dt} = \frac{\sigma_0 + \beta t}{E_0 M}\left[\frac{M}{M-N(t)}\right]^2 \dot N(t) + \frac{\beta}{E_0}\frac{M}{M-N(t)}. \tag{54}$$

which indicates that strain rate is governed by the interplay of damage rate (first term) and stress rate (second term). Combining equations (49), (53), and (54), we have:

$$\begin{aligned}\dot\varepsilon(t) &= \frac{E_0}{M(\sigma_0+\beta t)}\varepsilon(t)^2 \dot N(t) + \frac{\beta}{\sigma_0+\beta t}\varepsilon(t)\\ &= \frac{E_0 A^2}{M(\sigma_0+\beta t)}\left[1-\mathrm{sign}(m_\varepsilon)\left(\frac{t_c-t}{t_c}\right)^{m_\varepsilon}\right]^2 \dot N(t) + \frac{\beta A}{\sigma_0+\beta t}\left[1-\mathrm{sign}(m_\varepsilon)\left(\frac{t_c-t}{t_c}\right)^{m_\varepsilon}\right].\end{aligned} \tag{55}$$

Again, let us consider the regime where $t$ approaches $t_c$. If $\beta t_c \ll \sigma_0$, the first term dominates, given that the stress is approximately constant, and we obtain $p_\varepsilon + p_N \approx 2$ for $p_\varepsilon > 1$ ($m_\varepsilon < 0$), and $p_\varepsilon \approx p_N$ for $p_\varepsilon < 1$ ($m_\varepsilon > 0$), consistent with the constant stress case. If $\beta t_c \gg \sigma_0$, the stress increases significantly, yet the first term still dominates, as shown by the following derivation: (i) for $p_\varepsilon < 1$ (effective modulus diverges at $t_c$), from equation (54), we expect that the first term scales as $(t_c - t)^{p_N - 2}$ and the second term as $(t_c - t)^{p_N - 1}$, such that their ratio diverges as $t \to t_c$ (here, we used $M - N(t) = \int_t^{t_c}\dot N(\tau)d\tau \sim \int_t^{t_c}(t_c-\tau)^{-p_N}d\tau = \int_0^{t_c-t}(t_c-t-u)^{-p_N}du$ which scales as $(t_c-t)^{1-p_N}$ for $0 < p_N < 1$ and diverges for $p_N \geq 1$, with the latter physically irrelevant as the cumulative damage should remain finite at failure); (ii) for $p_\varepsilon > 1$ (effective modulus converges at $t_c$ so does the strain $\varepsilon$), the first equality in equation (55) indicates that the first term scales as $(t_c - t)^{-p_N}$, whereas the second term remains bounded, such that their ratio also diverges as $t \to t_c$. The dominance of the first term is also expected when $\beta t_c$ is comparable to $\sigma_0$. Therefore, the relationships among the different power law exponents given by equation (51), which were derived under constant stress conditions, are also expected to hold for linearly varying stress. While we have not examined more complex loading paths, these relationships may remain valid under broader forms of time-varying stress, provided the strain rate continues to be primarily controlled by the damage rate rather than by the stress rate. However, caution is warranted when generalizing to cases where the stress itself evolves over time with strong nonlinearity.

To conclude, we have developed an event-based cumulative damage modeling framework that captures the progression of geophysical systems towards catastrophic failure by focusing on sequences of discrete damage events. This approach reveals distinct finite-time singularities, including logarithmic and power law types, arising from the interplay between event timing and magnitude, and incorporates stochasticity to reflect natural variability. Through a mean-field model, we linked the power law exponents of different observables such as strain rate, event rate, and energy release rate to evolving system stiffness. These findings provide a simple unified theoretical picture for interpreting precursory signals across diverse geophysical systems and advance our understanding of the fundamental mechanisms driving catastrophic failure.

**Acknowledgments**

We gratefully acknowledge the financial support from the Norwegian Water Resources and Energy Directorate (NVE) through the project "Towards a Next-Generation Landslide Early Warning System". D.S. acknowledges partial support from the National Natural Science Foundation of China (Grant No. U2039202, T2350710802) and from the Shenzhen Science and Technology Innovation Commission (Grant No. GJHZ20210705141805017). During the preparation of this work the authors used ChatGPT in order to improve the readability and language of the manuscript. After using this tool/service, the authors reviewed and edited the content as needed and take full responsibility for the content of the published article.

**Open Research**

No data were used or generated in this study.